# Charge-cluster glass in an organic conductor

F. Kagawa[1,2†], T. Sato[1], K. Miyagawa[1], K. Kanoda[1], Y. Tokura[1,3], K. Kobayashi[4], R. Kumai[2,4], Y. Murakami[4]

*[1]Department of Applied Physics, University of Tokyo, Tokyo 113-8656, Japan*

*[2]CREST, Japan Science and Technology Agency (JST), Tokyo 102-0076, Japan*

*[3]RIKEN Center for Emergent Matter Science (CEMS), Wako 351-0198, Japan*

*[4]Condensed Matter Research Center (CMRC) and Photon Factory, Institute of Materials Structure Science, High Energy Accelerator Research Organization (KEK), Tsukuba, 305-0801, Japan*

† To whom correspondence should be addressed. E-mail: kagawa@ap.t.u-tokyo.ac.jp

Geometrically frustrated spin systems often do not exhibit long-range ordering, resulting in either quantum-mechanically disordered states, such as quantum spin liquids[1], or classically disordered states, such as spin ices[2,3] or spin glasses[4]. Geometric frustration may play a similar role in charge ordering[5,6], potentially leading to unconventional electronic states without long-range order; however, there are no previous experimental demonstrations of this phenomenon. Here, we show that a charge-cluster glass evolves upon cooling in the absence of long-range charge ordering for an organic conductor with a triangular lattice, $\theta$-(BEDT-TTF)$_2$RbZn(SCN)$_4$. A combination of time-resolved transport measurements and x-ray diffraction reveal that the charge-liquid phase has charge clusters that fluctuate extremely slowly (<10-100 Hz) and heterogeneously. Upon further cooling, the cluster dynamics freeze, and a charge-cluster glass is formed. Surprisingly, these observations correspond to recent ideas regarding the structural glass formation of supercooled liquids[7-10], indicating that a glass-forming charge liquid relates correlated-electron physics to glass physics in soft matter.

Wigner-type charge ordering (CO) is a phenomenon in which an equal number of charge-rich and charge-poor sites occupy a lattice such that rich-rich (or poor-poor) neighbouring pairs are avoided as much as possible. However, in a geometrically frustrated lattice, this constraint is insufficient to determine a specific CO among the various charge configurations, analogously to spin-frustrated systems (Fig. 1a); thus, the geometric frustration potentially undermines the tendency toward long-range CO, as was first suggested by Anderson[5]. As a result, exotic electronic states may be exhibited at low temperatures when long-range CO is avoided.

The material investigated in this study is the organic conductor $\theta$-(BEDT-TTF)$_2$RbZn(SCN)$_4$ (denoted $\theta$-RbZn), where BEDT-TTF (ET) denotes bis(ethylenedithio)tetrathiafulvalene[11]. The crystal structure consists of alternating layers of conducting ET molecules and insulating anions, and the ET molecules form a geometrically frustrated triangular lattice (Fig. 1b). The ET conduction band is hole-1/4-filled; thus, the charge-delocalized state (the "charge liquid" phase) is subject to CO instability[12]. However, the charge frustration created by the triangular lattice may prevent long-range ordering. Experimentally, a frustration-relaxing structural transition with the modulation $\mathbf{q_o} = (0, 0, 1/2)$ is shown to occur at ~200 K to stabilize a horizontal CO that matches $\mathbf{q_o}$ (ref. 13-15). This charge ordering is a strong first-order transition accompanied by a sudden increase in resistivity (Fig. 1c). When rapidly cooled (>5 K/min), the charge/structure order at 200 K vanishes, giving way to a charge liquid phase with a triangular lattice that is maintained at lower temperatures (Fig. 1c) (ref. 16,17). As there is no resistivity anomaly at 200 K during rapid cooling (Fig. 1c) and no trace of the $\mathbf{q_o}$ structural modulation above 200 K, even in the diffuse scattering[13] (see also Fig. 3a), the phase exhibited above 200 K is considered a continuation of the rapidly cooled phase. Interestingly, previous NMR measurements, which are a probe of local spin dynamics, implied the existence of slow charge dynamics on the order of kHz above 200 K (ref. 18). These slow dynamics may indicate that the charge liquid above 200 K is transforming into a classically disordered state, i.e., an electronic glass, but the relevance of this result to known glass formers is currently far from evident. To demonstrate possible electronic-glass-forming behaviour, several key concepts need to be tested, e.g., the temperature evolution of the slow charge dynamics, the spatial and dynamical heterogeneity, and the electronic-glass transition.

First, to directly detect the evolution of possible slow charge dynamics by determining their frequency, we implemented noise measurements, i.e., resistance fluctuation spectroscopy[19]. The typical resistance power spectrum density $S_R$ is shown in Fig. 2a. The global $S_R$ can be well fitted by $f^{-\alpha}$, with $\alpha \sim 0.8$-$0.9$ (weakly temperature dependent, $f$ denotes frequency). This fitting reflects the prevalent "$1/f$ noise", where the frequency exponent is not necessarily 1 (ref. 20). The origin of the ubiquitous "$1/f$ noise" was not identified for the present case and is beyond the scope of this study. Instead, we note that there is a finite deviation from the background "$1/f$ noise" over a certain frequency range (Fig. 2a). This deviation indicates that resistance fluctuations with a characteristic frequency are superimposed on the featureless "$1/f$ noise" (see Supplementary Information). For clarity, $f^{\alpha} \times S_R$ vs. $f$ is plotted for select temperatures in Fig. 2b, where the additional contribution appears as a broad peak rising out of a constant background (for this material, our methods are only applicable above 200 K; see Methods).

Upon quantifying the non-$1/f$ contributions, we found that the linewidth of the power spectrum density in the $f^{\alpha} \times S_R$ representation was broader than the expected linewidth for a single Lorentzian multiplied by $f^{\alpha}$ (Fig. 2c). To evaluate the linewidth, we introduced a hypothetical superposition of continuously distributed Lorentzians with high frequency $f_{c1}$ and low frequency $f_{c2}$ cutoffs (see Supplementary Information). We found that the spectrum shape is well reproduced using this scheme (Figs. 2b and 2c). The extracted values for $f_{c1}$ and $f_{c2}$ (Fig. 2d) reflect the fastest and slowest fluctuators, respectively, that are relevant to the non-$1/f$ contributions, from which we can estimate the centre frequency $f_0$ [defined as the geometric mean $(f_{c1} f_{c2})^{1/2}$] and the linewidth (defined as the ratio $f_{c1}/f_{c2}$). The temperature profiles of $f_0$ and $f_{c1}/f_{c2}$ are shown in Figs. 2e and 2f, respectively, and two features can be emphasized. First, $f_0$ slows down by several orders of magnitude as the temperature decreases; remarkably, just above the frustration-relaxing structural transition (approximately ~200 K), the lifetime of the long-lived fluctuator is less than 10 Hz (Fig. 2d). Second, the ratio $f_{c1}/f_{c2}$ noticeably increases, i.e., the dynamics become more heterogeneous, as the temperature decreases. These results are reproducible and are therefore considered to arise from the intrinsic nature of $\theta$-RbZn. We note that slow dynamics accompanied by dynamical heterogeneities are known to be key experimental properties for the vitrification process in supercooled normal liquids[21].

Some simulation results for supercooled normal liquids argued that medium-range "crystalline" clusters are critical for understanding the heterogeneous slow dynamics at a microscopic level[7-10], although the importance of these crystalline clusters is controversial[22-24], and there are still no clear experimental observations of crystalline clusters. To obtain further insight into the observed heterogeneous slow dynamics for the charge liquid, it is necessary to determine whether a type of "charge" cluster evolves with the slow dynamics. We conducted x-ray diffuse scattering measurements. A typical photographic oscillation image is shown in Fig. 3a. As previously reported[13], diffuse scattering characterized by $\boldsymbol{q_d}$ ~ ($\pm 1/3\ k$ $\pm 1/4$) is observed near the Bragg spots, where $k$ denotes negligible coherence between the ET layers. We note that this diffuse modulation, which is a clear indication of charge clusters comprised of a $3 \times 4$-period CO, cannot represent the nucleation of the horizontal CO with $\boldsymbol{q_o}$ [= (0 0 1/2)] below 200 K. The new feature shown in this study is the temperature profile of the charge clusters. In Figs. 3b and 3c, the evolution of both the intensity and the correlation length $\xi$ as a function of temperature can be observed. The value of $\xi$ is not short-ranged, but it is ~110 Å at 210 K, which corresponds to ~20 triangular spacings. This behaviour is essentially distinct from conventional critical phenomena, where $\xi$ diverges for a continuous transition. Based on a comparison of Figs. 3c and 2e, the growth of the slow dynamics and of $\xi$ appear well correlated, indicating that the charge clusters cause the heterogeneous slow dynamics. We also note that of the many different ideas regarding the liquid-glass transition in supercooled normal liquids, our experimental observations for the charge liquid agree phenomenologically with the proposals that consider the "crystalline" clusters to be the main source of heterogeneous slow dynamics[7-10].

All of the physical properties discussed so far indicate that the charge liquid phase above 200 K is transforming into a charge cluster glass. Thus, the last issue to be studied is the glass transition of the charge clusters, which is expected to occur only if the frustration-relaxing transition at 200 K is avoided by rapid cooling (Fig. 1c; see also Supplementary Information). A charge cluster glass transition is observed in the $\xi$ temperature profile measured during heating after rapid cooling (~90 K/min) (Fig. 3d). From 120 to 150 K, $\xi$ is temperature-insensitive with a significantly shorter length than expected (see also Fig. 3c), and no superlattice reflections attributable to a deformation of the triangular lattice are observed. This "frozen" metastable state with no long-range order is characteristic of a glassy state, indicating that a charge cluster glass is formed in the charge-frustrated triangular lattice. Upon further heating, $\xi$ increases sharply to the expected value at $T_g$ ~160-170 K and becomes temperature-dependent for temperatures above $T_g$. This behaviour demonstrates that the charge liquid nature is recovered above $T_g$, and $T_g$ can therefore be regarded as the charge cluster glass transition temperature. Interestingly, previous a.c. conductance measurements reported a broad anomaly at approximately 167 K during rapid cooling[16], which we think is another indication of the charge cluster glass transition. The observations of (i) frozen charge clusters, (ii) glass transition behaviour, and

(iii) glass-forming characteristics exhibited by the temporal and spatial properties of the charge liquid phase demonstrate that a charge liquid confined to a charge-frustrated triangular lattice is subject to charge cluster vitrification in an organic system, which is generally considered clean.

Inhomogeneous electronic states are often found in strongly correlated electron systems that contain dopants[25-28], such as manganites and high-transition temperature cuprates, in which randomly distributed dopants are considered to play a major role in the emergent inhomogeneity[29,30]. Conversely, the organic conductor used in this study is nominally dopant-free and, when slowly cooled, exhibits a well-defined CO with no phase mixture[13-15], indicating that the microscopic mechanism of the charge cluster glass formation is essentially distinct from the mechanism of known electronic inhomogeneities.

**Methods Summary**

**Crystal growth**

Single crystals of $\theta$-(BEDT-TTF)$_2$RbZn(SCN)$_4$ were synthesized by the galvanostatic anodic oxidation of BEDT-TTF (50 mg) in a N$_2$ atmosphere, in accordance with ref. 11.

**Noise measurements**

The noise measurements were conducted using the conventional four-terminal d.c. method. A steady current was supplied from a low-noise voltage source, and the voltage between the voltage-probing electrodes was fed into a spectrum analyzer (Agilent 35670A) after amplifying it with a low-noise preamplifier. A large load resistor was used to eliminate the resistance fluctuation effect at the contacts. For the material used in this study, this method cannot be applied below 200 K because the centre frequency $f_0$ is too slow; the relaxation from the supercooled charge liquid to the more stable horizontal CO occurs before the measurements are completed.

**Diffraction scattering experiments**

The diffraction of a θ-RbZn single crystal (0.4×0.3×2.5 mm$^3$) was measured using a Rigaku DSC imaging plate diffractometer and Si(111) monochromatized synchrotron radiation x-rays (λ = 0.689 Å) at the BL-8B beamline of the Photon Factory (PF) at the High Energy Accelerator Research Organization (KEK). The correlation length of the charge clusters is defined as ξ = ($l/\pi$)FWHM$^{-1}$, where FWHM denotes the full width at half maximum, and $l$ is the unit cell length along the direction studied.

**Supplementary Information** accompanies the paper.

**Acknowledgements**     We thank H. Tanaka, Y. Kohsaka, K. Kuroki, H. Seo, M. Watanabe, R. Kurita, and M. Imada for fruitful discussions. This work was partially supported by JSPS through the "Funding Program for World-Leading Innovative R&D on Science and Technology (FIRST Program)" and by JSPS KAKENHI (Grant Nos. 24684020, 20110002, and 24654101).

**Author Contributions**     F.K. and T.S. conducted the noise measurements. T.S., K. Kobayashi., R.K., K.M., and F.K. conducted the x-ray measurements. K.M. grew the single crystals used for the study. F.K. and K. Kanoda planned and headed the project. F.K. wrote the letter with assistance from Y.T., K. Kanoda, and Y.M.

**Author Information**     The authors declare no competing financial interests. Correspondences and requests for materials should be addressed to F.K. (kagawa@ap.t.u-tokyo.ac.jp).

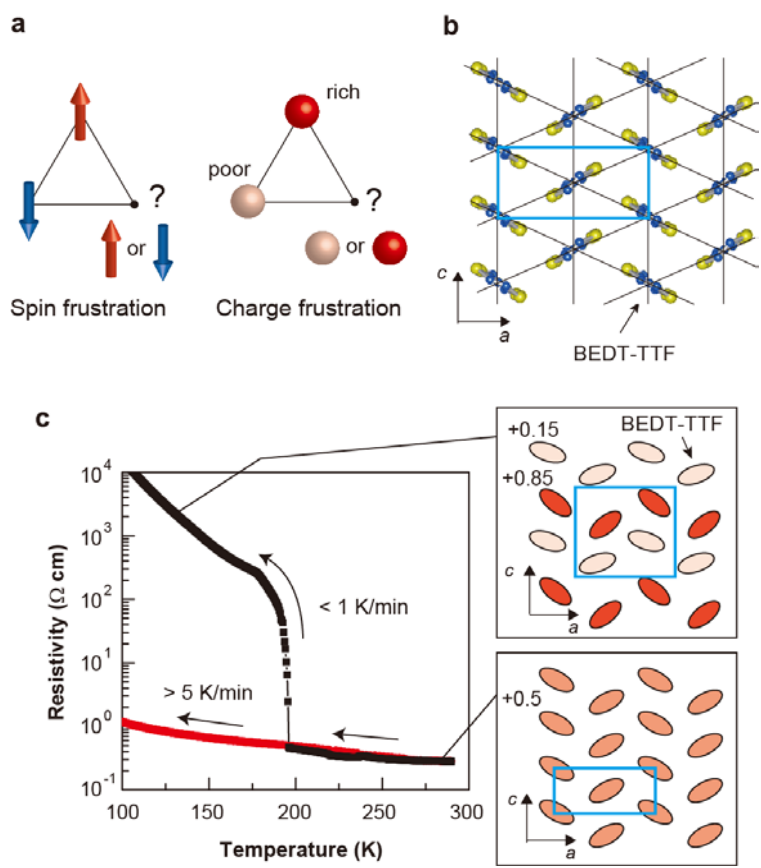

**Figure 1 │Charge frustration and crystal structure of θ-(BEDT-TTF)₂RbZn(SCN)₄.**

**a**, A schematic illustrating the analogy between spin frustration and charge frustration. **b**, The structure of the BEDT-TTF layer. **c**, The temperature dependence of the resistivity during cooling for different temperature-sweeping rates. The insets schematically illustrate the crystal structures of the high-temperature phase (lower inset) and the low-temperature phase (upper inset). In the upper inset, the charge-ordering pattern is also shown. To emphasize the two-fold structural modulation of the *c*-axis, the upper inset is depicted in an exaggerated manner. In each panel, the unit cell is indicated by a blue rectangle.

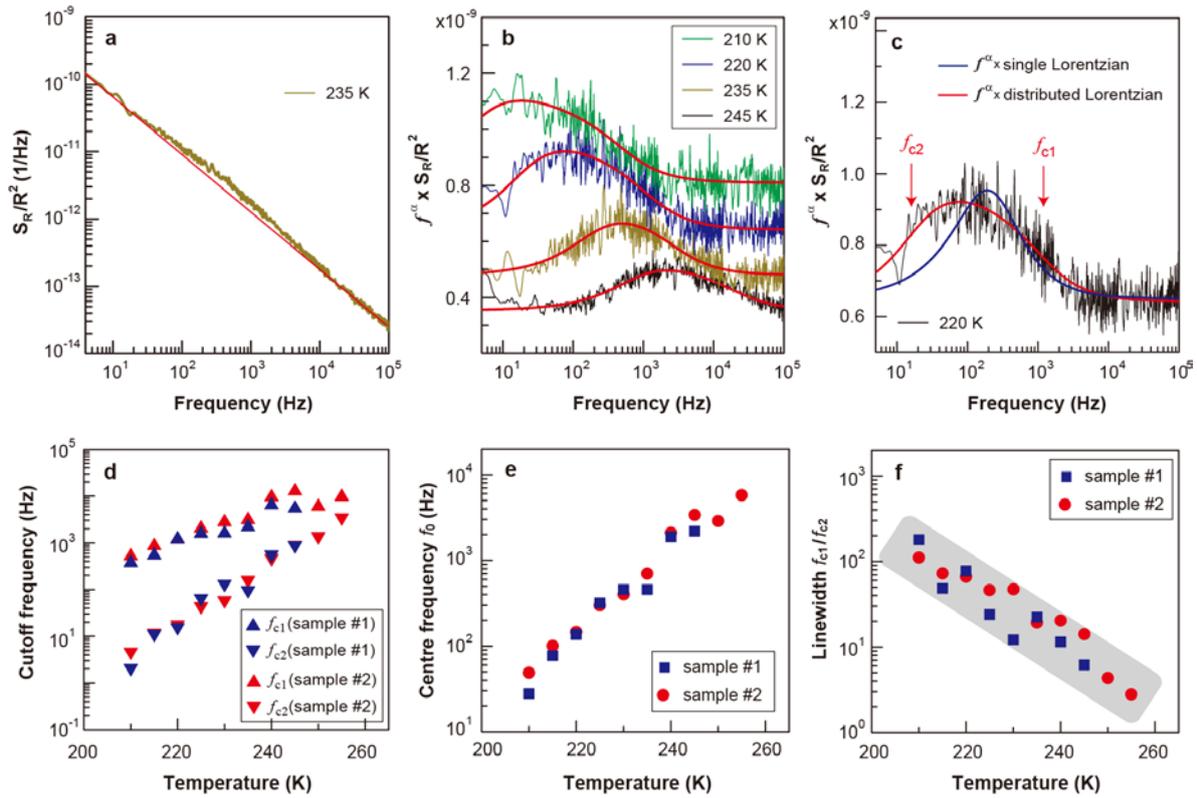

**Figure 2 │ Resistance fluctuations under a steady electrical current. a**, A typical resistance power spectrum density $S_R$ normalized by the resistance squared $R^2$. **b**, Power spectra densities for various temperatures with $f'' \times S_R/R^2$ representations. The red lines are fits to the distributed Lorentzian model (see Supplementary Information). **c**, A comparison of the $S_R/R^2$ characterizations using different schemes. The notations $f_{c1}$ and $f_{c2}$ represent the high- and low-frequency cutoffs, respectively, in the distributed Lorentzian model. **d**, Temperature profiles of the fitting parameters $f_{c1}$ and $f_{c2}$. **e, f,** Slowing of the centre frequency (**e**) and concomitant growth of the dynamical heterogeneity (**f**). The shaded area in **f** is intended as a guide for the eyes.

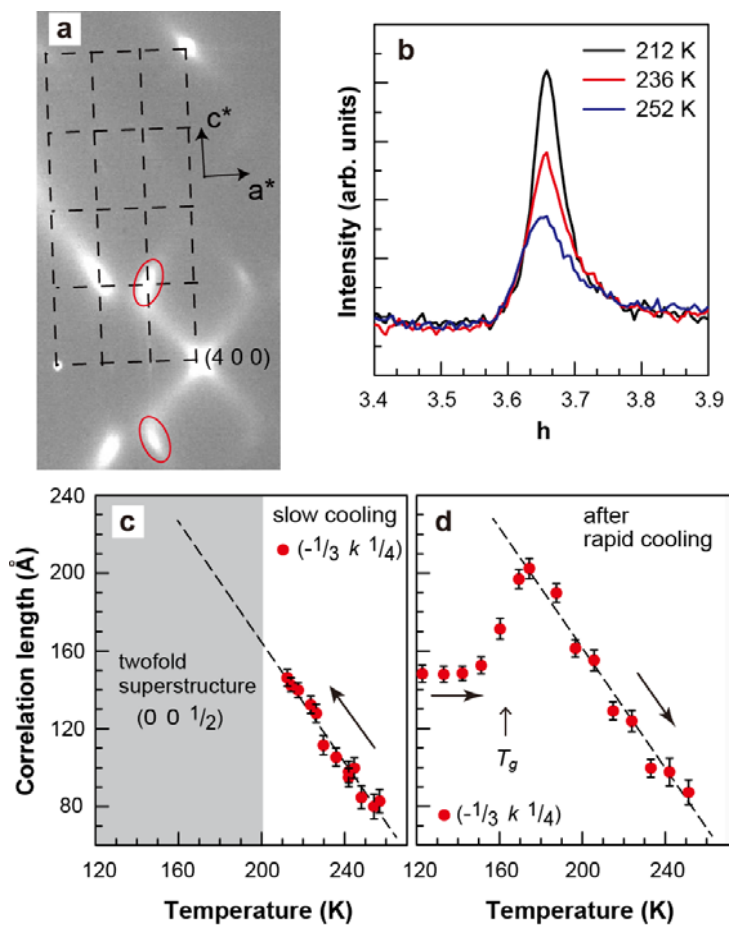

**Figure 3 │ Spatial correlation of the charge clusters investigated using x-rays. a**, Photographic oscillation image of the $a^*$-$c^*$ plane at 225 K. Diffuse rods characterized by $\boldsymbol{q} \sim (\pm 1/3\ k\ \pm 1/4)$ are observed near the Bragg reflections (some are indicated by ellipsoids in the image). **b**, Line profile of $\boldsymbol{q} \sim (4\text{-}1/3\ k\ 0\text{+}1/4)$ along the $3a^*$-$4c^*$ direction. **c, d,** Temperature dependence of the charge cluster correlation length $\xi$ during slow cooling (**c**) and during heating after rapid cooling to 120 K (**d**). The value of $\xi$ is estimated along the $(\text{-}1/3\ k\ 1/4)$ direction on the $(4\text{-}1/3\ k\ 0\text{+}1/4)$ diffuse rod. The broken lines in **c** and **d** are drawn as guides for the eyes. The error bars represent the numerical ambiguity of the fitting.

Supplementary Information for

# Charge-cluster glass in an organic conductor

F. Kagawa[1,2†], T. Sato[1], K. Miyagawa[1], K. Kanoda[1], Y. Tokura[1,3], K. Kobayashi[4], R. Kumai[2,4], Y. Murakami[4]

[1]*Department of Applied Physics, University of Tokyo, Tokyo 113-8656, Japan*

[2]*CREST, Japan Science and Technology Agency (JST), Tokyo 102-0076, Japan*

[3]*RIKEN Center for Emergent Matter Science (CEMS), Wako 351-0198, Japan*

[4]*Condensed Matter Research Center (CMRC) and Photon Factory, Institute of Materials Structure Science, High Energy Accelerator Research Organization (KEK), Tsukuba, 305-0801, Japan*

† To whom correspondence should be addressed. E-mail: kagawa@ap.t.u-tokyo.ac.jp

## Single Lorentzian model

To characterize the resistance power spectrum density (PSD) observed in the experiments, it is useful to refer to the fluctuating two-level system that has the same energies (Fig. S1a). The voltage PSD produced by switching between two states takes the form of a Lorentzian (S1):

$$S_R(f) = A\frac{\tau_c}{1+(2\pi f\tau_c)^2},\qquad(1)$$

where $A$ and $\tau_c$ represent the amplitude of the resistance fluctuation and the characteristic half-life (Fig. S1a), respectively. The frequency profile $S_R(f)$ is shown in Fig. S1b, where the characteristic (corner) frequency, $f_c$, is given by $f_c = 1/(2\pi\tau_c)$.

In a typical PSD, the ubiquitous "$1/f$ noise" often coexists with a Lorentzian. Hence, the total PSD is given as

$$S_R(f) = A\frac{\tau_c}{1+(2\pi f\tau_c)^2}+\frac{B}{f^\alpha},\qquad(2)$$

where $B$ represents the amplitude of the "$1/f$ noise". The total PSD when $\alpha = 1$ is shown in Fig. S1c, where the Lorentzian contribution is seen as a deviation from the $1/f$ background. In the $f\times S_R$ representation of the total PSD (Fig. S1d), the $1/f$ noise and the Lorentzian contribution appear as a constant background and as a peak rising out of this background, respectively. The peak frequency corresponds to $f_c$.

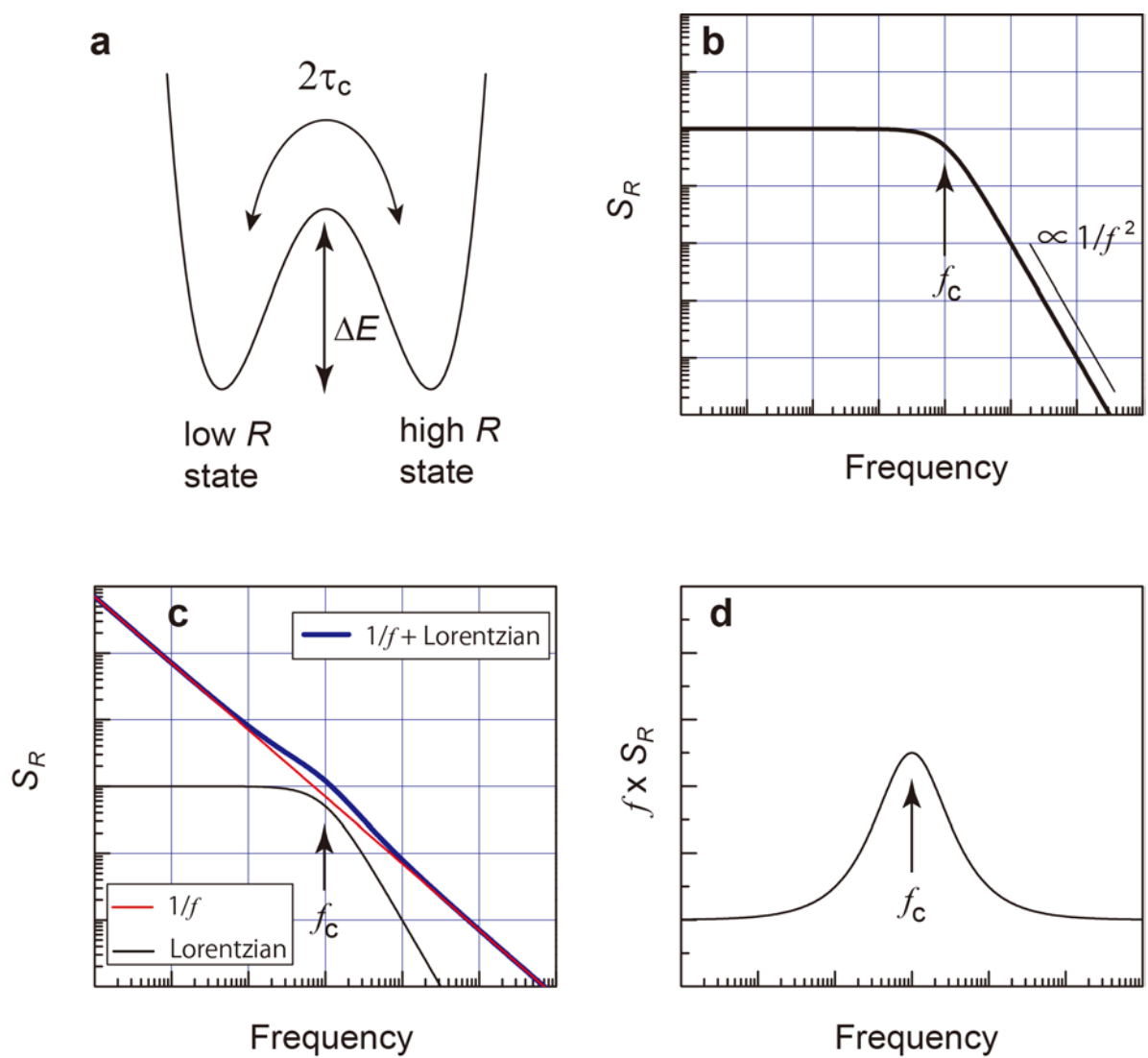

**Figure S1.** Simulated results of the power spectrum density $S_R$ associated with a single Lorentzian model. **a,** Model of a double-well potential with a potential barrier $\Delta E$ and characteristic lifetime $2\tau_c$. **b,** Power spectrum density $S_R$ of a single Lorentzian. "$f_c$" denotes $1/(2\pi\tau_c)$. **c,** The power spectrum density when a single Lorentzian coexists with the $1/f$ noise. The experimentally observable power spectrum density is the sum of the $1/f$ contribution and the Lorentzian contribution. **d,** The $f \times S_R$ representation of the power spectrum density shown in **c**.

## Distributed Lorentzian model with cutoff frequencies

As discussed in the main text, we found that the observed non-$1/f$ contributions exhibit broadened linewidth; therefore, they could not be well reproduced using the single Lorentzian model (see Fig. 2c in the main text). Therefore, we have to consider a distribution in the potential barrier $\Delta E$ or, equivalently, in $\tau_c$. For simplicity, we make two assumptions as follows: First, the following form of the density of states, $g(\Delta E)$, is postulated (Fig. S2a):

$$g(\Delta E) = \begin{cases} \text{const.} & (\Delta E_1 \leq \Delta E \leq \Delta E_2) \\ 0 & (\text{otherwise}) \end{cases}. \tag{3}$$

Because $\tau_c$ is given by $\tau_c \sim \exp(\Delta E/k_B T)$, $g(\Delta E)$ can be rewritten as a function of $\tau_c$ (Fig. S2b):

$$G(\tau_c) \propto \begin{cases} 1/\tau_c & (\tau_{c1} \leq \tau_c \leq \tau_{c2}) \\ 0 & (\text{otherwise}) \end{cases}, \tag{4}$$

where $\tau_{c1}$ and $\tau_{c2}$ denote the shorter and longer cutoff lifetimes, respectively. Second, we assume that all fluctuators have the same resistance fluctuation amplitudes.

Based on these assumptions, the power spectrum density of our hypothetical distributed Lorentzian model can be calculated analytically as follows:

$$S_R(f) = A \int_{\tau_{c1}}^{\tau_{c2}} G(\tau_c) \frac{\tau_c}{1 + (2\pi f \tau_c)^2} d\tau_c,$$

$$= A' \int_{\tau_{c1}}^{\tau_{c2}} \frac{1}{1 + (2\pi f \tau_c)^2} d\tau_c$$

$$= \frac{A'}{2\pi f} \left[ \tan^{-1}(2\pi f \tau_{c2}) - \tan^{-1}(2\pi f \tau_{c1}) \right] \tag{5}$$

This frequency profile is shown in Fig. S2c, where $f_{c2}$ and $f_{c1}$ denote $1/2\pi\tau_{c2}$ and $1/2\pi\tau_{c1}$, respectively. As discussed above, the $1/f^\alpha$ noise is superimposed onto Eq. (5) in actual experiments. To mimic the experimental results of θ-RbZn, we consider the case of α = 0.86. Figure S2d shows the PSD for the sum of Eq. (5) and $1/f^{0.86}$. Figure S2e shows the $f^{0.86} \times S_R$ representation of the PSD shown in Fig. S2d.

Although Eq. (5) is derived on the basis of some assumptions, we found that the observed data are well reproduced by this scheme (see Figs. 2b and 2c in the main text), thus allowing us to evaluate the temperature dependence of $f_{c1}$ and $f_{c2}$ (Fig. 2d in the main text).

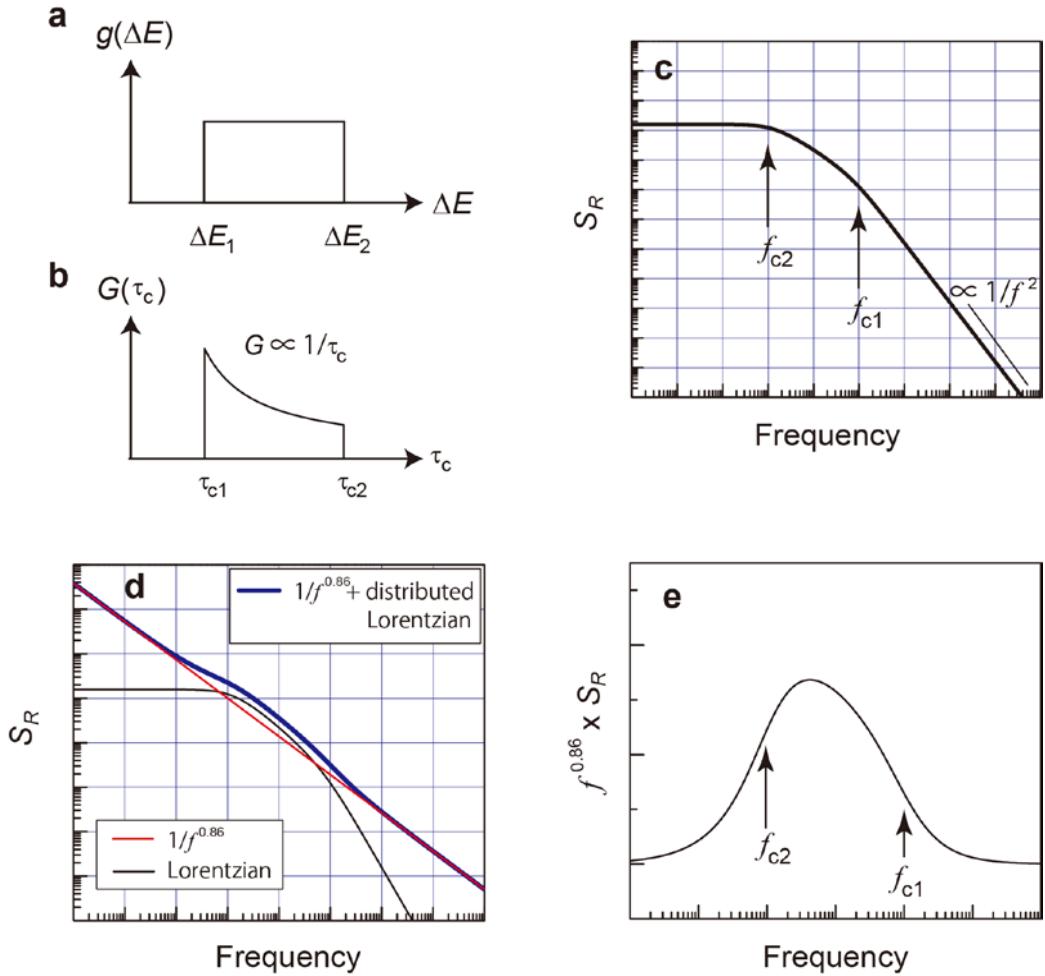

**Figure S2.** Simulated results of the power spectrum density $S_R$ associated with our distributed Lorentzian model. **a** and **b,** The postulated density of states of the Lorentzian as a function of the potential barrier (**a**) and lifetime (**b**). **c,** The power spectrum density $S_R$ of our distributed Lorentzian model. "$f_{c1}$" and "$f_{c2}$" denote the characteristic frequencies of the fastest and slowest fluctuators, respectively. **d,** The power spectrum density when the $1/f^{\alpha}$ noise with $\alpha = 0.86$ coexists with our distributed Lorentzian. The experimentally observable power spectrum density is a sum of the $1/f^{0.86}$ contribution and the distributed Lorentzian contribution. **e,** The $f^{0.86} \times S_R$ representation of the power spectrum density shown in **d**.

**Kinetic avoidance of horizontal charge ordering at ~200 K**

The symmetry of the charge clusters that evolve in the charge-liquid phase (i.e., >200 K) differs from the symmetry of the horizontal charge order that emerges at temperatures of less than 200 K (under slow cooling). Therefore, it is plausible that the charge clusters are unfavorable to the formation of the horizontal charge order. At 200 K, the medium-range charge clusters are still dynamic (on the order of Hz, see Fig. 2e in the main text), i.e., they can be temporally created or annihilated. Therefore, the charge clusters would behave as removal disorders when slowly cooled, compared with the characteristic lifetime of the clusters, but as quenched disorders when rapidly cooled. This dual nature likely results in the cooling-rate-dependent bifurcation of the formation and avoidance of the horizontal charge order.

**Temperature dependence of the correlation length of charge clusters**

In Ref. S2, the numerical simulations clearly demonstrate that the correlation length $\xi$ of medium-range crystalline order evolves with a power-law divergence as a function of temperature toward the ideal glass-transition point $T_0$, while we observed the linear temperature dependence of $\xi$ (Figs. 3c and 3d). Here we discuss the origin of difference.

We note that $T_0$ is not the same as the experimentally determined glass-transition point $T_g$, which should be higher than $T_0$ by definition. As is mentioned in the main text, the $T_g$ is not a thermodynamic transition but of dynamic origin, which arises from the competition between the system relaxation time and the experimental cooling rate; therefore $T_g$ generally depends on the cooling rate. Contrastingly, $T_0$ (if any) is a thermodynamic transition and hence independent of cooling rate, although it is practically impossible to hold a supercooled state down to $T_0$ in the laboratory time scale because of the diverging relaxation time (i.e., crystallization occurs before the system reaches $T_0$). The reason why the power-low behavior was not observed in Fig. 3d is that the transition-like behaviour at 160-170 K is the experimental glass transition $T_g$, not the ideal glass transition $T_0$. The $T_0$ and critical regime (if any) should be located at temperatures much lower than $T_g$, practically inaccessible in experiments.

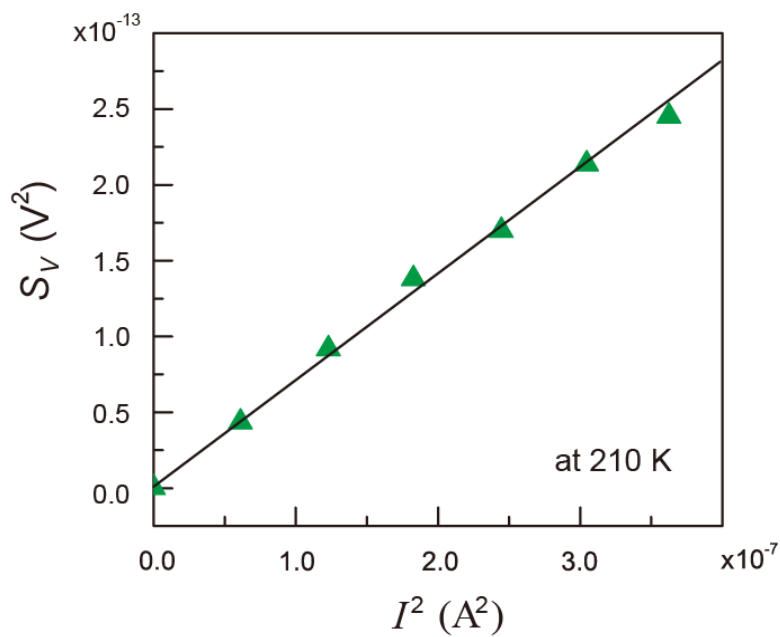

**Figure S3.** Scaling behavior of $S_V$ with the square of the applied current. The integrated noise power from 160 Hz to 190 Hz is shown.

# Supplementary References